\documentclass{article}
\usepackage{icrctc07}

\newcommand{\gray}[1]{$\gamma$-ray{#1}}

\newcommand{\pubjournal}[6] {#1, #2 {\bf #3}, #4 (#5).}

\newcommand{\icrc}{{\it Int. Cosmic Ray Conf.}}

\newcommand{\aap}{{\it Astron. Astrophys.}}
\newcommand{\apj}{{\it ApJ}}

\newcommand{\EB}{EGRB}

\title{A Monte Carlo Study of the Irreducible Background in the EGRET Instrument}
\shorttitle{EGRET Irreducible Monte Carlo Study}

\authors{T. A. Porter, W. B. Atwood, B. Baughman, and R. P. Johnson}
\shortauthors{Porter et al.}
\afiliations{Santa Cruz Institute for Particle Physics, University of California, Santa Cruz, CA 95064, U.S.A.}
\email{tporter@scipp.ucsc.edu}

\abstract{The diffuse extragalactic \gray{} background (\EB) has been 
derived by various groups from observations by the Energetic Gamma Ray 
Experiment Telescope (EGRET) instrument on the Compton Gamma Ray Observatory 
(CGRO). 
The derived EGRB consists of \gray{s} that may come from 
astrophysical components, such as from unresolved extragalactic point 
sources (blazars, normal galaxies, etc.), true extragalactic diffuse emission, 
misattributed diffuse signals 
from the Galaxy and other celestial sources, 
and an irreducible instrumental background due to \gray{s} produced by 
cosmic-ray (CR) interactions in the EGRET instrument.
Using the Gamma Ray Large Area Space Telescope (GLAST) simulation and 
reconstruction software, we have investigated the magnitude of the irreducible
instrumental background in the GLAST Large Area Telescope (LAT). 
We re-scale our results to the EGRET
and present preliminary results of our study and its effect on current 
estimates of the EGRB.}

\begin{document}
\maketitle
\section{Introduction}

The diffuse extragalactic \gray{} background (\EB) is a
superposition of all unresolved sources and true diffuse 
high energy \gray{} emission in the Universe.  
The \EB\ is a
weak component which is difficult to disentangle from the intense Galactic
foreground. 
The isotropic, presumably extragalactic component of the
diffuse \gray{} flux was first discovered by the SAS-2 satellite 
\cite{Thompson1982}.
Analysis of the data from the EGRET instrument on CGRO appears to 
confirm this discovery \cite{Sreekumar1998}. 

The extraction of the \EB\ is difficult because its derivation relies 
on modelling foregrounds that are uncertain to some degree, as well as 
a good understanding of the instrumental background.
Extensive work has been done \cite{Sreekumar1998} to derive the
spectrum of the \EB\ based on the EGRET data.  
The relation of modelled
Galactic diffuse emission to total measured diffuse emission was used
to determine the \EB\, as the extrapolation to zero Galactic
contribution.  
A new detailed model of the Galactic diffuse emission \cite{Strong2004a}
lead to a new estimate of the \EB\ which is lower and steeper
than found by \cite{Sreekumar1998}.
Analysis of the same data by Grenier et al.~\cite{Grenier2005} found similar 
results; they make the important point that
the overall intensity and spectrum depend within 50\% on the 
choice of foreground model.

Understanding of the instrumental background is also crucial
for extraction of the \EB{}.
Gamma-ray telescopes, such as the EGRET and the upcoming GLAST-LAT, employ 
a sensitive anti-coincidence shield (ACS) to veto charged particles entering
the instrument field of view (FoV).
Surrounding the ACS there is additional material, such as the 
thermal blanket and micrometeor shield.
Charged particles interacting in this inert material can produce neutral 
secondaries and not trigger the ACS.
Similarly, if a charged particle interacts in the scintillator in the ACS
without causing a veto, the neutral 
secondaries will enter the instrument in the FoV.
In either case, the secondaries contaminate the celestial signal and are an 
irreducible background that is a systematic uncertainty in determining
the level of the \EB{}. 




In this paper, we report on a study of the irreducible background
in the GLAST-LAT, and its application to estimate 
the systematic uncertainty in the \EB\ derived from the EGRET data.

\section{Monte Carlo Simulations and Analysis}

The GLAST-LAT is a pair-conversion telescope in which the tracker-converter 
uses silicon microstrip technology to track the electron-positron pairs 
resulting from \gray{} conversion in thin tungsten foils.
A cesium iodide calorimeter below the tracker is used to measure the 
\gray{} energy, and the tracker is surrounded on the other 5 sides by plastic
scintillators forming the ACS for charged-particle rejection.

As part of the pre-launch analysis, simulation studies of data collected 
by the LAT are performed.
These include a complete detector simulation with 
realistic orbit and attitude profile, full CR background model,
and a detailed model of the \gray{} sky including 
time variable sources. 
The resulting simulation data are pushed through an analysis chain which 
includes direction and energy reconstruction, background rejection and 
event classification algorithms allowing the identification of 
well-reconstructed \gray{} events.

\begin{figure*}
\begin{center}
\includegraphics[width=0.65\textwidth]{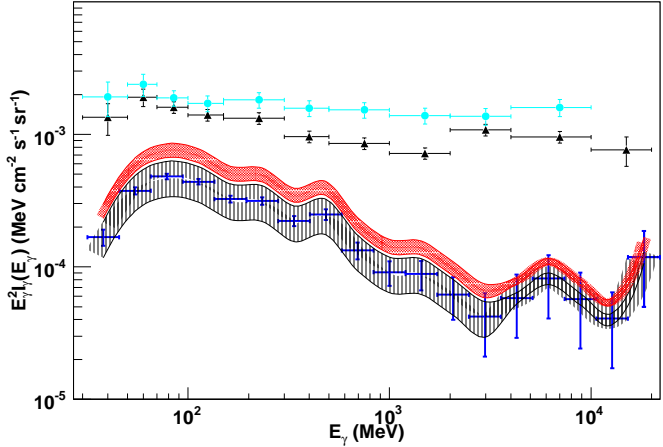}
\end{center}
\caption{Extragalactic \gray{} and irreducible background in 
the GLAST-LAT based on our Monte Carlo study.
Data points: cyan-circle from \cite{Sreekumar1998}; 
black-triangle from \cite{Strong2004a}.
Blue points with error bars: irreducible intensity from our analysis.
Black-lines show the uncertainty on the irreducible intensity due to 
uncertainties in the charged particle flux model.
Black-hatched region shows the combined uncertainty from the charged 
particle flux model and hadronic physics modelling.
Red-shaded region shows the hatched region re-scaled 
to the nominal CGRO orbital altitude of 450 km.}\label{fig1}
\end{figure*}

The presence of so-called irreducible \gray{} events became apparent upon 
scanning the residual background events after their statistical 
rejection in the analysis.
Gamma-rays in this event class are produced in the inert material located 
outside of the sensitive portion of the ACS scintillation tiles (this 
includes $\sim1$ mm of the plastic scintillator since it is possible for 
particles to interact in the scintillator producing all neutral secondaries 
without sufficient light production to cause a reliable veto).
The incident charged particles responsible for these \gray{s} were:
positrons (60\%), protons producing $\pi^0$s (30\%), and 
electrons/positrons producing \gray{s} via Bremsstrahlung (10\%).
Our handscan of the residual events resulted in a sample of 751 irreducible
background events.
Using these events, a live-time of $2\times10^4$ seconds 
and an effective area-solid angle product of $2.2\times10^4$ cm$^2$ sr,
the irreducible \gray{} intensity was computed in the LAT.
Figure~\ref{fig1} shows the result of our analysis for the irreducible 
\gray{} component in the GLAST-LAT.
The error bars are statistical only.


There is a systematic uncertainty on the irreducible component due to the
uncertainty in the incident charged particle fluxes.
The details of the particle flux model are outlined in \cite{Ormes2007}.
Albedo electrons, positrons, and protons are included, as well as 
Galactic CRs.
Included in the albedo component are splash particles produced by CR
interactions with the atmosphere and re-entrant particles trapped by the 
Earth's magnetic field.
The albedo electron/positron component is dominant 
below $\sim 400-500$ MeV while the 
albedo proton component is the main contributor up to $\sim 3$ GeV; above 
this Galactic CRs are the dominant component of the charged particle flux.
The major 
uncertainties associated with orbit-averaged components of the flux model are:

$\bullet$ $\pm50$\% for albedo electrons/positrons; 
$\pm30$\% for albedo electrons/positrons with energies $\geq 30$ MeV

$\bullet$ $\pm30$\% for albedo protons

$\bullet$ $\pm10$\% for the Galactic CR component.




Furthermore, there is an additional significant systematic error due to the 
hadronic physics modelling for protons producing $\pi^0$s without causing 
a veto in the ACS.
Hadronic interactions are less well-modelled compared with electromagnetic 
interactions and we have taken a conservative additional 
20\% uncertainty for the proton induced irreducible component.
We combine these uncertainties and show them as the hatched band in 
Fig.~\ref{fig1}.

To re-scale our results for the EGRET we need to account for the difference
in orbital altitude between the GLAST and the CGRO.
The GLAST-LAT study was done for an orbital altitude 565 km, whereas the
the nominal orbital altitude for the CGRO when deployed was 450 km.
However, the CGRO's orbital altitude decreased by $\sim 100$ km requiring 
subsequent reboosts to regain the initial deployed altitude.
The uncertainty in the charged particle flux model for lower orbital 
altitudes is: 

$\bullet$ $20$\% increase in the albedo component as orbital altitude 
decreases from 615-430 km

$\bullet$ $10$\% decrease in the Galactic CR component as orbital altitude
decreases from 615-430 km.

\noindent
We adopt a simple $20$\% increase in the albedo component and $10$\% decrease
in the Galactic CR component.
We do not consider the decrease in orbital altitude that the CGRO experienced; 
this would add further systematic uncertainty.
The irreducible intensity from our analysis of the GLAST-LAT 
re-scaled to the nominal CGRO orbital altitude is shown in Fig.~\ref{fig1}
as the shaded region.

The difference between the GLAST-LAT and the EGRET inert material audit must 
also be taken into account.
However, it is the most uncertain step of a process that already involves
considerable uncertainties.
For the LAT, the micrometeor shield and thermal blanket have a total column
density of 0.38 g cm$^{-2}$, with the $\sim 1$ mm of inert scintillator 
contributing a further 0.15 g cm$^{-2}$ giving a total of 0.53 g cm$^{-2}$.
For the EGRET, the micrometeor shield, thermal blanket, and light shield amount
to a column density of 0.20 g cm$^{-2}$.
For the EGRET ACS we can only estimate the column density of inert material
by examining the veto threshold energy.
The veto threshold energy was measured during the EGRET calibration but was
known to have significant systematic errors.
Thus, we are only able to give a range of the charged particle 
penetration depth before which a veto would be triggered in the ACS.
From examining internal EGRET documents \cite{Kanbach2007}, we find that 
the penetration depth can range from as little as 0.15 mm up to 2.5 mm for
the apex of the ACS.
This yields a range of total column densities of inert material in the EGRET
of 0.215 g cm$^{-2}$ to 0.45 g cm$^{-2}$.
A simple scaling of the derived irreducible intensity by the relative 
column densities between the LAT and the EGRET is not possible due to the
uncertainties associated with the charged particle rejection from hadronic
interactions in the inert material.
Therefore, we do not attempt to further re-scale the derived irreducible
intensity, but simply mention that there will be a further systematic error
from the uncertainty in the amount of inert material.


\section{Discussion}

We have made a study of the irreducible background in the GLAST-LAT and 
have extended this, 
using simple re-scalings of the particle flux model and an estimate of the
EGRET inert material audit, to estimate the irreducible background in the EGRET.

From this analysis, the importance of accurately determining the irreducible 
component is clear.
To enable this, additional information derived from the MC truth has been 
added to the GLAST-LAT data sets.
This includes the location that the incident particle intersects the surface 
of the LAT, the energy it deposits in the ACS tile it hits, and a count of 
the number of hits registering in the silicon strip tracker caused by 
particles other than electrons/positrons and photons.   
With this information the signature of an irreducible event becomes 
its direction within the FoV, its intersection point within the fiducial 
volume of the tracker, and the lack of any hits caused by anything other than
electrons/positrons, and photons.

We have attempted to re-scale the results of our analysis to estimate 
the irreducible background in the EGRET.
Our estimate is not exact given the considerable 
uncertainties associated with the 
charged particle flux model, the variation in the orbital altitude of CGRO over 
its mission, the amount of inert material in the EGRET, 
and the hadronic physics in the Monte Carlo model.
However, it is a non-negligible fraction of current estimates of the \EB{}.

The \EB{}, as inferred from the EGRET data, is affected by large systematic
errors.
The data are strongly affected by the Galactic foreground subtraction, as well
as uncertainties in the irreducible instrumental background.
Taken together, these uncertainties imply that current estimates of the \EB\
should realistically be viewed as upper limits only.

\section{Acknowledgements}
This work was done under US Department of Energy grant DE-FG02-04ER41286
and NASA contract PY-1775.


\begin{thebibliography}{9}

\bibitem{Grenier2005}
  \pubjournal{Grenier, I.~A., Casandjian, J.~M., \& Terrier, R.}
             {\icrc}{4}{13}{2005}




\bibitem{Kanbach2007}
  Kanbach, G., {\it private communication}.

\bibitem{Ormes2007}
  Ormes, J. F., et al., 
  to appear in {\it Proc. 1$^{\rm st}$ Int. GLAST Symp.}
  (Stanford, Feb. 5-8, 2007), eds. Ritz, S., Michelson, P.~F., \& Meegan, C.,
  {\it AIP Conf. Proc.}, astro-ph/0704.0462.  
 
\bibitem{Sreekumar1998}
  \pubjournal{Sreekumar, P., et al.}{\apj}{494}{523}{1998}


\bibitem{Strong2004a}
  \pubjournal{Strong, A.~W., Moskalenko, I.~V., \& Reimer, O.}
             {\apj}{613}{956}{2004}

\bibitem{Thompson1982}
  \pubjournal{Thompson, D. J., \& Fichtel, C. E.}{\aap}{109}{352}{1982}

\end{thebibliography}
\end{document}